\documentclass[12pt]{iopart}
\usepackage{graphicx}

\begin{document}

\title[LISA Long-arm interferometry]{LISA Long-arm Interferometry}

\author{James Ira Thorpe}

\address{NASA/GSFC, Greenbelt, MD 20771, USA}

\ead{james.i.thorpe@nasa.gov}
\begin{abstract}

The Laser Interferometer Space Antenna (LISA) will observe gravitational
radiation in the milliHertz band by measuring picometer-level fluctuations
in the distance between drag-free proof masses over baselines of approximately
$5\times10^{9}\,\mbox{m}$. The measurement over each baseline will
be divided into three parts: two 'short-arm' measurements between
the proof masses and a fiducial point on their respective spacecraft,
and a 'long-arm' measurement between fiducial points on separate spacecraft.
This work focuses on the technical challenges associated with these
long-arm measurements and the techniques that have been developed
to overcome them. 

\end{abstract}



\section{Introduction}

Gravitational waves from distant astrophysical sources can be detected by monitoring the deviation
between two or more inertial test particles separated over a long baseline. The Laser Interferometer
Space Antenna \cite{Bender_98} (LISA) is a realization of such a
detector in the $10^{-4}\,\mbox{Hz}\leq f\leq10^{-1}\,\mbox{Hz}$ frequency band,
a regime rich in astrophysical sources \cite{Hughes_06}. The instrument
consists of three separate sciencecraft (SC) in a triangular formation
approximately $5\,\mbox{Gm}$ on a side. By monitoring the distance
between reference points on separate SC with a precision of $\sim10^{-11}\,\mbox{m}$,
gravitational wave signals with strains on the order of $\sim10^{-21}$
can be observed.

The test particles for the LISA measurement are the gravitational
reference sensors (GRSs), each consisting of a $4\,\mbox{kg}$ cube
of Au-Pt alloy known as a {}``proof-mass'' that floats freely inside
an electrode housing. A control system maintains the gap between the
proof mass and the housing by actuating thrusters on the SC, a scheme
known as drag-free control. 

\begin{figure}
\begin{centering}
\includegraphics[width=6cm]{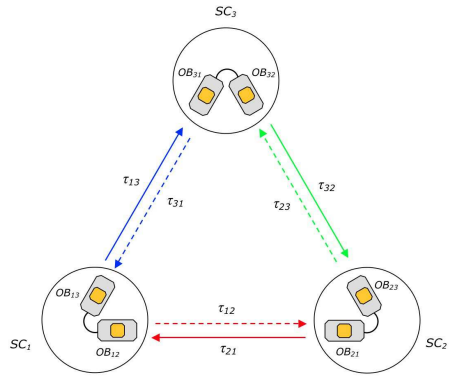}
\par\end{centering}
\caption{\label{fig:constellation}The LISA constellation consists of three
sciencecraft (SC), each containing two optical benches (OBs). The
long-arm measurement consists of measuring displacements along each
of the six one-way links, characterized by $\tau_{ij}$, the light
travel time from $SC_{i}$ to $SC_{j}$.}
\end{figure}

Each SC contains two GRS units, one oriented towards each of the other
SC, as shown in \ref{fig:constellation}. The goal of the interferometric
measurement system (IMS) is to monitor the distance fluctuations along
each of the six one-way links between the proof-masses on separate
SC. For convenience, each one-way measurement is broken into three
sub-measurements: two short-arm measurements of the distance between
each proof-mass and a reference point on its host SC and a long-arm
measurement between reference points on the two separate SC. 

This work focuses on the long-arm measurement and relevant aspects
of the subsystems involved in making it. Section \ref{sec:Subsystems}
describes the subsystems that are involved in the long-arm
interferometry while section \ref{sec:Interferometery} describes how the
problem of laser phase noise is addressed.
As with other recent review articles \cite{Shaddock_08,Jennrich_09},
the intent is to give an overview of work done by the community of
researchers contributing to LISA interferometry, of which the author
is but one member. 

\section{Relevant Subsystems\label{sec:Subsystems}}

There are several subsystems involved in making the long-arm measurement.
In this section, an overview of these subsystems is presented with
an emphasis on their role in long-arm interferometry. The subsystems
considered are the laser subsystem, the telescope, the optical bench,
the pointing mechanisms, and the phase measurement subsystem. It is
worth noting that the subsystems discussed here are the ones with
the greatest difference between LISA and its technology demonstrator
mission, LISA Pathfinder \cite{McNamara_08,Armano_09} (LPF), a single-SC
mission which does not include a long-baseline measurement.

\subsection{Laser subsystem\label{sub:Laser}}

The laser subsystem provides the light source for the interferometric
measurements. The Nd:YAG laser must deliver a nominal optical power
of $\sim1\,\mbox{W}$ at $1.064\,\mu\mbox{m}$ to the transmitting
telescope. The center frequency of the laser must be tunable, both
to enable heterodyning between multiple laser sources and to allow
the laser frequency/phase to be actively controlled to suppress the
intrinsic noise of the laser. A final requirement for the laser system is the
ability to phase modulate the laser at high frequency (several GHz), a capability that is used for the clock noise correction discussed in section \ref{sub:PMS}.

The current baseline design for the LISA laser is a master-oscillator
power-amplified (MOPA) architecture. The master oscillator, in the
form of a non-planar ring oscillator (NPRO) such as the one that will
fly on LPF or a fiber laser\cite{Numata_10}, produces a low-power
beam which can be frequency tuned. Light from the master laser is
modulated in a waveguide phase modulator and used to seed a fiber
amplifier which produces the final output power. This design requires
that neither the phase modulator nor power amplifier introduce phase
noise at the $\sim1\,\mu\mbox{cycle}$ level. Tests of both components
are underway and preliminary results indicate that this architecture
can meet the requirements \cite{Trobs_09}.

\subsection{Telescope\label{sub:Telescope}}

The function of the telescope subsystem is to exchange beams between
optical benches on separate SC. The primary concern is the amount
of power which can be collected by the receiving SC, as this sets
the shot noise limit for the displacement measurement. Assuming an
end-of-life transmitted power of $0.5\,\mbox{W}$, the received power
will be as low as $100\,\mbox{pW}$, corresponding to a
shot noise limit on displacement of $\sim10^{-11}\,\mbox{m}/\sqrt{\mbox{Hz}}$\cite{Jennrich_09}.
All other IMS noise sources are designed to be below this shot noise
limit.

A second concern for the telescope is its mechanical stability, which
directly enters the long-baseline measurement. To keep mechanical
fluctuations of the telescope below shot noise, the pathlength through the telescope must
be stable to $\sim1\,\mbox{pm}/\sqrt{\mbox{Hz}}$. If this level of
stability cannot be met, an alternative is to measure pathlength
fluctuations in the telescope with an auxiliary interferometer and
correct for them in post-processing, an arrangement
known as an optical truss.

Other concerns for the telescope are the quality of the wavefront
delivered to the far SC and the generation of scattered light. Deviations
from an ideally spherical wavefront at the receiving SC will cause
angular jitter in the transmitting SC to produce a phase shift in
the received beam that would be interpreted as a displacement. Light
from the transmitting SC may scatter off of the telescope and interfere
with the received light, generating an additional heterodyne signal
with a phase that is dependent on the stability of the scattering
path.

\subsection{Optical Bench\label{sub:Bench}}

\begin{figure}
\begin{centering}
\includegraphics[width=8cm]{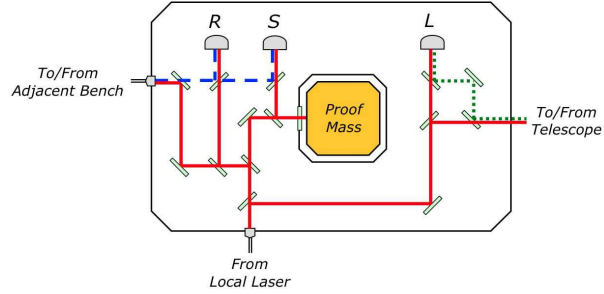}
\par\end{centering}

\caption{\label{fig:opticalBench}Schematic of the LISA optical bench showing
three primary interferometric measurements (reference, $R_{ij}$,
short-arm, $S_{ij}$, and long-arm, $L_{ij}$) made using the three
beams: from the local laser (solid), from the adjacent bench via the
inter-bench fiber (dashed), and from the far SC via the telescope
(dotted).}

\end{figure}

The function of the optical bench is to interfere the various beams
needed to make the displacement measurements. Figure \ref{fig:opticalBench}
shows a schematic (not a realistic layout) of the bench, with three
beams entering the bench. The local beam enters via an optical fiber
from the laser subsystem, a beam from the adjacent bench enters via
the {}``back-link'' fiber that connects the two benches on each SC, and a beam
from the far SC enters via the telescope. The local laser is also
transmitted to the far SC via the telescope.

These three beams are used to make three heterodyne interferences.
The reference measurement ($R$), combines the local beam with the
adjacent beam to measure the relative phase noise of the two lasers
as well as the phase noise in the back-link fiber. The short-arm measurement
($S$) is identical to the reference measurement except that the local
beam is reflected off of the proof mass, adding the projected motion
of the SC to the measurement. The long-arm measurement ($L$) combines
the local beam with the received beam, measuring laser phase noise,
motion of both SC, shot noise, and gravitational wave strain. In each of these signals, the heterodyne phase
is dominated by the intrinsic phase noise of the interfering beams. The key
to LISA interferometry is to measure these signals with high fidelity
and then combine them in ways that cancel laser phase noise and other
obscuring terms in order to extract the desired quantities. 

As with the telescope, the pathlength stability of the optical bench
is also critical. The LISA optical bench will be constructed using
a technique known as hydroxide-catalysis bonding that has been applied to the LPF optical bench\cite{Elliffe_05,Killow_06}. In this technique,
the individual optics are bonded to an ultra-stable glass substrate,
yielding a bench with a stability approximately equivalent to that
of a monolithic structure.

\subsection{Constellation Design and Pointing Mechanisms\label{sub:Constellation}}

\begin{figure}
\begin{centering}
\includegraphics[width=12cm]{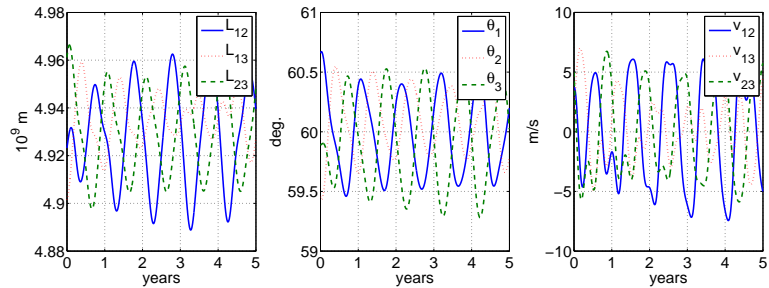}
\par\end{centering}

\caption{\label{fig:orbits}Arm-lengths, interior angles, and line-of-sight
velocities for one realization of the LISA constellation. This particular
realization is designed to minimize the line-of-sight velocities.}

\end{figure}

The LISA constellation is not
actively maintained through station-keeping; each SC is inserted into
the appropriate orbit and the constellation geometry evolves along
with those orbits. Perturbations from the planets and other bodies
cause the orbits to oscillate with an annual period as well as experience secular degradation. Orbits are designed using optimization
techniques \cite{Hughes_08} that seek to constrain some property
of the constellation. For example, the orbits in Figure \ref{fig:orbits} minimize the relative line-of-sight velocities
and resulting Doppler shifts in the long-arm heterodyne signals.

The $\sim1^{\circ}$ variations in the interior angle of the LISA
constellation are larger than the field of view of the telescope and
hence require a mechanism to maintain pointing to both far SC simultaneously.
The optical assembly tracking mechanism (OATM) provides this function
by pivoting the payload assembly (GRS, optical bench, \& telescope)
to track one of the far SC while the other one is tracked by adjusting
the attitude of the SC. The closed-loop pointing error of the OATM
must have a spectral density less than $\sim1\,\mbox{nrad}/\mbox{\ensuremath{\sqrt{\mbox{Hz}}}}$
in the LISA band. Achieving this stability and accuracy over a $1^{\circ}$
dynamic range requires a multi-stage mechanism or an actuator with
high accuracy and large dynamic range, such as a piezo inchworm.

An additional mechanism is needed to compensate for the relative motion
between the SC. Due to the light travel time between the SC
($\tau_{ij}\sim17\,\mbox{s}$), it is necessary to point the transmitted
beam ahead of the position of the received beam. This introduces
an angle between the transmitted and received beams on the order of
$\Delta v/c$, where $\Delta v$ is the relative velocity between
the SC, that must be corrected prior to interfering the beams. For
the orbital solution in Figure \ref{fig:orbits}, the angle has a
component in the constellation plane with a nominal value of $\sim1.6\,\mu\mbox{rad}$
and variations of $\sim60\,\mbox{nrad}$ as well as a component normal
to the constellation plane that is approximately a mean-zero sinusoid
with an amplitude of $\sim3\,\mu\mbox{rad}$. A mechanism known as
the point-ahead actuator%
\footnote{If the actuator is implemented on the received beam, it is sometimes
referred to as a look-behind actuator.%
} (PAA) corrects for these angles, maintaining alignment between the
local and received beam. Unlike the OATM, the PAA is in the interferometric
path and consequently must keep any added displacement noise below
the $\sim1\,\mbox{pm}/\sqrt{\mbox{Hz}}$ level.

\subsection{Phase Measurement Subsystem\label{sub:PMS}}

The function of the phase measurement subsystem (PMS) is to measure
the phase of the heterodyne signals on the six optical benches with
sufficient fidelity that they can be combined to measure desired quantities
such as SC motion and GW signals. The primary challenge for the PMS
is one of dynamic range: the laser phase noise has a spectral density
in excess of 1=$1\,\mbox{MHz}/\sqrt{\mbox{Hz}}$ in the LISA band
whereas the desired noise floor for detecting gravitational wave signals
is on the order of $\sim1\,\mu\mbox{cycle}/\sqrt{\mbox{Hz}}$. An
additional complication in the $L$ measurement is the variation of
the heterodyne frequency caused by the time-varying Doppler between
the SC (see section \ref{sub:Constellation}).

The PMS can be divided into an analog chain, which converts the optical
beat note into a digital signal, and a digital processor, which extracts
the phase of that signal. The analog chain consists of the photoreceiver,
amplifiers, and analog-to-digital converters. Each beat note is measured
using a quadrant detector, which allows the angle between the interfering
beams to be measured by differencing the measured phase from different
quadrants. The digital portion of the PMS consists of a digital oscillator
that is phase-locked to the heterodyne signal. The phase of the heterodyne
signal is extracted by measuring the frequency of the phase-locked
oscillator as well as the phase error in the loop. Implementations
of the PLL-type PMS \cite{Shaddock_06,Bykov_09} have demonstrated
the ability to measure LISA-like signals with a dynamic range of better
than $\sim10^{14}$ in the LISA band, more than sufficient for LISA
requirements.

The phase measurements made on each SC will be contaminated by phase
noise of the ultra-stable oscillator (USO) used provide the time reference
on-board the SC. Ordinarily, the USO phase noise is not a concern
as it is common to all measurements made on the SC. In LISA, however,
measurements from multiple SC are combined, each with their own independent
USO noise. To address this issue, the relative phase noise of each
pair of USOs is measured by transmitting a phase modulation tone along
each of the inter-SC links\cite{Klipstein_06}. Recording the phase of this tone allows
the relative USO clock noise to be removed in post-processing.

\section{Mitigation of Laser Phase Noise\label{sec:Interferometery}}

One of the main challenges for LISA interferometry is mitigation of
laser phase noise that couples into the long-arm measurement. In a
classic equal-arm Michelson interferometer, the phase noise of the
light source does not couple into the displacement measurement because
it is common in both interfering beams. In the LISA long-arm interferometer,
this is not the case for two reasons: there are two independent light
sources and there is significant difference in the arm-length. At
a Fourier frequency of $1\,\mbox{mHz}$, the phase noise spectral
density of the free-running LISA lasers is expected to be on the order
of $5\times10^{9}\,\mbox{cycles}/\sqrt{\mbox{Hz}}$, more than 14
orders of magnitude above the roughly $10\,\mu\mbox{cycle}$ shot
noise limit. The following sections describe how this noise is suppressed
in the measurement, allowing LISA to operate at the shot noise limit.

\subsection{Time-Delay Interferometry\label{sub:TDI}}

The goal of Time Delay Interferometry\cite{Tinto_99,Shaddock_04b}
(TDI) is to recover the insensitivity to phase noise that is present
in an equal-arm Michelson interferometer. The concept is to combine
the phase measurements of the individual links in post processing
in order to synthesize an equal-arm interferometer. To do this requires
some knowledge of the constellation geometry as well as sufficiently
high fidelity in the phase measurements that the cancellation will
span the many orders of magnitude needed to reach the target sensitivity. 

The ability of TDI to suppress laser frequency noise is limited by
a number of effects, the relative importance of which can be expressed
using a suppression limit factor defined as the ratio between the
input phase noise and the phase noise in the TDI channel. Table \ref{tab:TDI_Limits}
lists several limiting effects that have been studied. Arm-length
knowledge refers to errors in the assumed values of the light travel
times that are used to form the delays in the TDI combinations. Algorithm
errors result from limitations in the ability to correct for higher-order
terms in the dynamics of the constellation (relative velocity, acceleration,
etc. between the SC). Interpolation errors are limits on the accuracy
of determining the value of the phase meter signal at a particular
time from a series of evenly-spaced points \cite{Shaddock_04b}. Analog
chain errors refer to effects of dispersion and non-linearity in PMS
analog chain. Scattered light errors refer to parasitic heterodyne
signals that are generated by stray beams, also known as small vector
noise.

\begin{table}
\centering{}\caption{\label{tab:TDI_Limits}Phase noise suppression limits for TDI as a
function of Fourier frequency $f$ (Adapted form \cite{Shaddock_09})}
\begin{tabular}{|c|c|c|}
\hline 
Effect & Assumption & Suppression Limit\tabularnewline
\hline
\hline 
Arm-length knowledge &  ranging error of $\delta x$ & $2.4\times10^{10}\times\left(1\,\mbox{m}/\delta x\right)\times\left(1\,\mbox{mHz}/f\right)$\tabularnewline
\hline 
Algorithm Errors & velocity-correcting TDI & $2\times10^{12}\times\left(1\,\mbox{mHz}/f\right)$\tabularnewline
\hline 
Interpolation Errors & $21\,\mbox{s}$ kernel, $3\,\mbox{Hz}$ sampling & $3.2\times10^{15}\times\left(1\,\mbox{mHz}/f\right)^{2}$\tabularnewline
\hline 
Analog Chain Errors & Laboratory measurements & $5\times10^{10}\times\left(1\,\mbox{mHz}/f\right)$\tabularnewline
\hline 
PMS noise floor & Laboratory measurements & $10^{14}\times(1\,\mbox{mHz}/f)^{2}$\tabularnewline
\hline 
Scattered light & amplitude of $2\times10^{-5}$  & $1.5\times10^{17}\times(1\,\mbox{mHz}/f)$\tabularnewline
\hline
\end{tabular}
\end{table}

The limiting effect for TDI is arm-length knowledge, which can be obtained form a
number of sources. In rough order of accuracy they are SC ephemeris
models, Doppler tracking of the SC form ground, and active ranging
on-board the SC. The baseline ranging system for LISA consists of
a pseudo-random noise (PRN) code stream added as a phase-modulation
sideband to the clock transfer tones. Additional channels in the PMS
measure the phase of the received signal and pass it to the ranging
processor, where it is tracked using a delay-locked loop.
The current estimate of the ranging error for such a system is $\delta x\sim1\,\mbox{m}$ \cite{Esteban_09}.

The LISA error budget allocates $2\,\mbox{pm}/\sqrt{\mbox{Hz}}\times\sqrt{1+\left(3\,\mbox{mHz}/f\right)^{4}}$
to residual laser phase noise, where $f$ is the Fourier frequency.
Assuming ranging-limited TDI, this sets a limit on the maximum allowable
laser frequency noise \emph{prior }to TDI, \begin{equation}
\tilde{\nu}_{pre-TDI}(f)=282\,\mbox{Hz}/\sqrt{\mbox{Hz}}\times\left(\frac{1\,\mbox{m}}{\delta x}\right)\times\sqrt{1+\left(\frac{3\,\mbox{mHz}}{f}\right)^{4}}.\label{eq:nu_req}\end{equation}
The estimated spectrum for the unstabilized LISA laser is $30\,\mbox{MHz}/\sqrt{\mbox{Hz}}\times\left(1\,\mbox{mHz}/f\right)$.
This leaves a gap of approximately six orders of magnitude between
the interferometry requirement and the capability of ranging-limited
TDI with $1\,\mbox{m}$ arm length knowledge. Consequently the LISA
lasers must be actively stabilized. 

\subsection{Constellation Phase Locking\label{sub:PLL}}

Active frequency stabilization of the LISA lasers is simplified through
the use of phase-lock loops (PLLs). One SC is designated as the master
SC and one of its lasers is designated as the master laser. All other
lasers in the constellation are phase locked to the master laser which
is then stabilized to a frequency reference. An advantage
of PLLs is the ability to add an arbitrary frequency offset to the
slave lasers. This extra degree of freedom can be used to ensure that
all heterodyne frequencies remain in the range of the PMS as the orbital
Doppler frequencies change (see section \ref{sub:Constellation}).
It also avoids the problem of having to match the absolute frequencies
of multiple independent frequency references to within the PMS bandwidth.

It should be noted that the application of PLLs does not effect the
TDI algorithms. The cancellation of laser phase noise in the TDI combinations
does not assume any particular correlation between the phase noises
from individual lasers, nor is it adversely effected if such correlations
exist.

\subsection{Arm-locking\label{sub:Arm-locking}}

One option for actively stabilizing the frequency of the master laser
is to use the LISA constellation as a frequency reference, a technique
referred to as arm-locking. The original proposal for arm-locking,
known as single arm-locking\cite{Sheard_03}, showed that the phase
of the long-arm heterodyne signal could be used to estimate the laser
phase noise. The difficulty with this approach is that this signal
is insensitive to phase noise at $f_{n}\equiv n/\tau_{RT,j}$ where $\tau_{RT}$
is the round-trip light travel time in the arm and $n=1,2,3,...$.
These sensor nulls lead to noise enhancement spikes in the closed-loop
system at the $f_{n}$ frequencies, which populate the LISA measurement
band. In addition, the slope of the loop gain for $f>f_{1}$ is limited
by the phase response of the sensor. Despite these drawbacks, single-arm
locking is able to provide significant frequency noise suppression
for $f\neq f_{n}$ and has been demonstrated in a number of hardware
models\cite{Marin_05,Thorpe_05,Wand_09,Sheard_05}.

Sensitivity to phase noise at $f=f_{n}$ can be recovered by utilizing
the adjacent LISA arm \cite{Sutton_08}, which generally has a different
value for $\tau_{RT}$ and hence a different set of $f_{n}$. With
two arms to provide the information, all frequencies are covered other
than those that are multiples of the inverse difference in the round
trip times, $f_{n}^{\prime}\equiv n/\Delta\tau_{RT}$ where $\Delta\tau_{RT}\equiv\left|\tau_{RT,1}-\tau_{RT,2}\right|$.
At these frequencies, the null points of the two individual arms overlap
and the same issues of instability arise. However, because $\Delta\tau_{RT}/\tau_{RT}\sim0.01$,
the $f_{n}^{\prime}$ frequencies lie above the LISA measurement band. 

There is some design freedom in determining how best to combine the
information from the two long-arm measurements. The goal is to generate
a sensor with a flat, broadband response to laser phase noise and
minimal coupling of phase noise from other sources, a problem that
can be approached using optimization techniques\cite{Maghami_09}
or traditional control design methods. The current baseline arm-locking
sensor is the \emph{modified dual arm-locking} sensor, described by
McKenzie, et al. \cite{McKenzie_09}. 

\begin{figure}
\begin{centering}
\includegraphics[width=10cm]{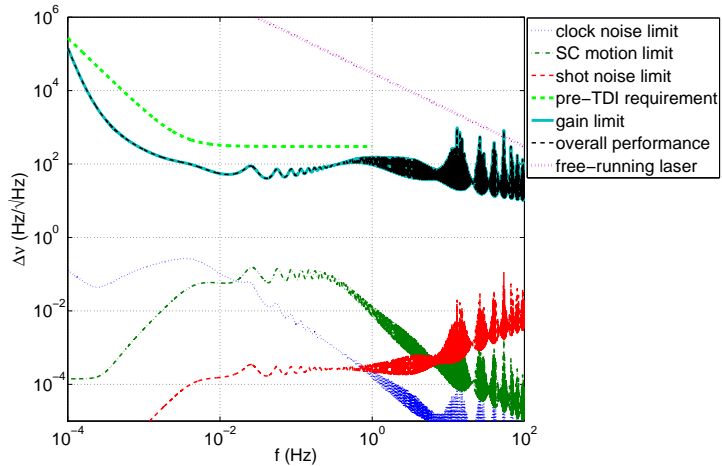}
\par\end{centering}

\caption{\label{fig:MDALS}Performance of a modified dual arm locking system
\cite{McKenzie_09} assuming free-running lasers and an arm-length
mismatch of $\Delta\tau_{RT}=0.037\,\mbox{s}$, the minimum value
for the orbits in Figure \ref{fig:orbits}. }

\end{figure}

Figure \ref{fig:MDALS} shows the performance of a system based on
this sensor and a complimentary controller design assuming a free-running
frequency noise of $30\,\mbox{kHz}/\sqrt{\mbox{Hz}}$ for the master
laser. The curves are plotted assuming $\Delta\tau_{RT}=0.037\,\mbox{s}$,
the minimum value for the orbits in Figure \ref{fig:orbits} assuming
that the master SC can be switched at any time to maximize $\Delta\tau_{RT}$.
The curve labeled 'pre-TDI requirement' is the maximum allowable frequency
noise at the input to TDI, assuming $1\,\mbox{m}$ ranging accuracy
(see section \ref{sub:TDI}). 

If a situation were to arise where one of the long-baseline links
was not functional, it would not be possible to switch the master
SC and $\Delta\tau_{RT}$ would occasionally reach zero. McKenzie estimates that this system would still meet LISA requirements
except during periods of minimum $\Delta\tau_{RT}$ lasting approximately
$30\,\mbox{min}$ and occurring twice per year. 

In addition to the gain and noise limitations to arm-locking described
above, there is an important limitation associated with the dynamic
range of the laser frequency. The PMS measures the phase of the long-arm 
heterodyne signal relative to a reference signal at the expected heterodyne
frequency. Finite knowledge of the inter-SC Doppler shift introduces a mismatch in these frequencies, which is manifested in the PMS output by a ramping phase term.. As the arm-locking
controller attempts to correct for this term, it pulls the laser
frequency, risking bringing
the master laser, or one of the other phase-locked lasers, into an
unstable operating regime. The problem of laser pulling can be mitigated
by improving estimates of the Doppler frequency and AC-coupling the
arm-locking controller to reduce the effects of Doppler errors.

\subsection{Local Stabilization\label{sub:Local-Stabilization}}

Another option is to stabilize the master laser to a frequency reference on-board the SC.
This is the approach taken by LISA Pathfinder, which uses a Mach-Zender
(MZ) interferometer with a deliberate pathlength difference to measure
and correct for frequency fluctuations\cite{Steier_09}. The performance
of such a system is limited by the pathlength stability of the MZ,
which is integrated into the optical bench, and the noise floor of
the PMS. The expected performance is approximately $1\,\mbox{kHz}/\sqrt{\mbox{Hz}}\times\sqrt{1+(3\,\mbox{mHz}/f)^{4}}$.
While this performance is not sufficient to reach the pre-TDI requirement,
the MZ can serve as a pre-stabilization stage for arm-locking, reducing
the requirements on arm-locking and increasing overall system margin.

The best-performing standard frequency references in the LISA band
are high-Finesse optical cavities\cite{Drever_83}. Mueller, et al.
\cite{Mueller_05} demonstrated performance of better than $30\,\mbox{Hz}/\sqrt{\mbox{Hz}}\times\sqrt{1+(3\,\mbox{mHz}/f)^{4}}$
in the LISA band, capable of meeting the pre-TDI requirement even
if the ranging accuracy is relaxed to $10\,\mbox{m}$.

In order to use an optical cavity as a pre-stabilization stage for
arm-locking, it is necessary to provide a way to adjust the center
frequency of the laser while still maintaining the stability provided
by the cavity. The preferred method for doing this is offset sideband
locking\cite{Thorpe_08}, which modifies the standard Pound-Drever-Hall
readout scheme to allow the laser frequency to have an adjustable
offset form the cavity resonance. This avoids the problems with corrupting
the intrinsic stability of the reference that arise when an adjustable
element such as a piezeo-electric actuator is added to the cavity
itself \cite{Conti_03}.

\subsection{Reference Designs}

In this section we consider four reference
designs for laser phase noise mitigation that meet LISA requirements. These designs
were proposed in a whitepaper \cite{Shaddock_09} prepared by the
LISA Frequency Control Working Group, an ad-hoc assembly of frequency
control researchers from government agencies, universities, and industry
who are actively working on LISA.

TDI with $1\,\mbox{m}$ ranging accuracy is common to all designs, setting the maximum pre-TDI frequency noise limit as $282\,\mbox{Hz}/\sqrt{\mbox{Hz}}\times\sqrt{1+(3\,\mbox{mHz}/f)^{4}}$.
All lasers are assumed to be phase locked to a single master laser,
as described in section \ref{sub:PLL}. In designs where arm-locking
is included, it is implemented using the modified dual arm-locking
as described by McKenzie, et al.\cite{McKenzie_09}. Arm-locking performance
is computed assuming a minimum round-trip difference of $\Delta\tau_{RT}=0.037\,\mbox{s}$,
corresponding to the minimum difference in the orbits described in
Figure \ref{fig:orbits}, assuming all inter-SC links are available.

\begin{figure}
\begin{centering}
\includegraphics[width=10cm]{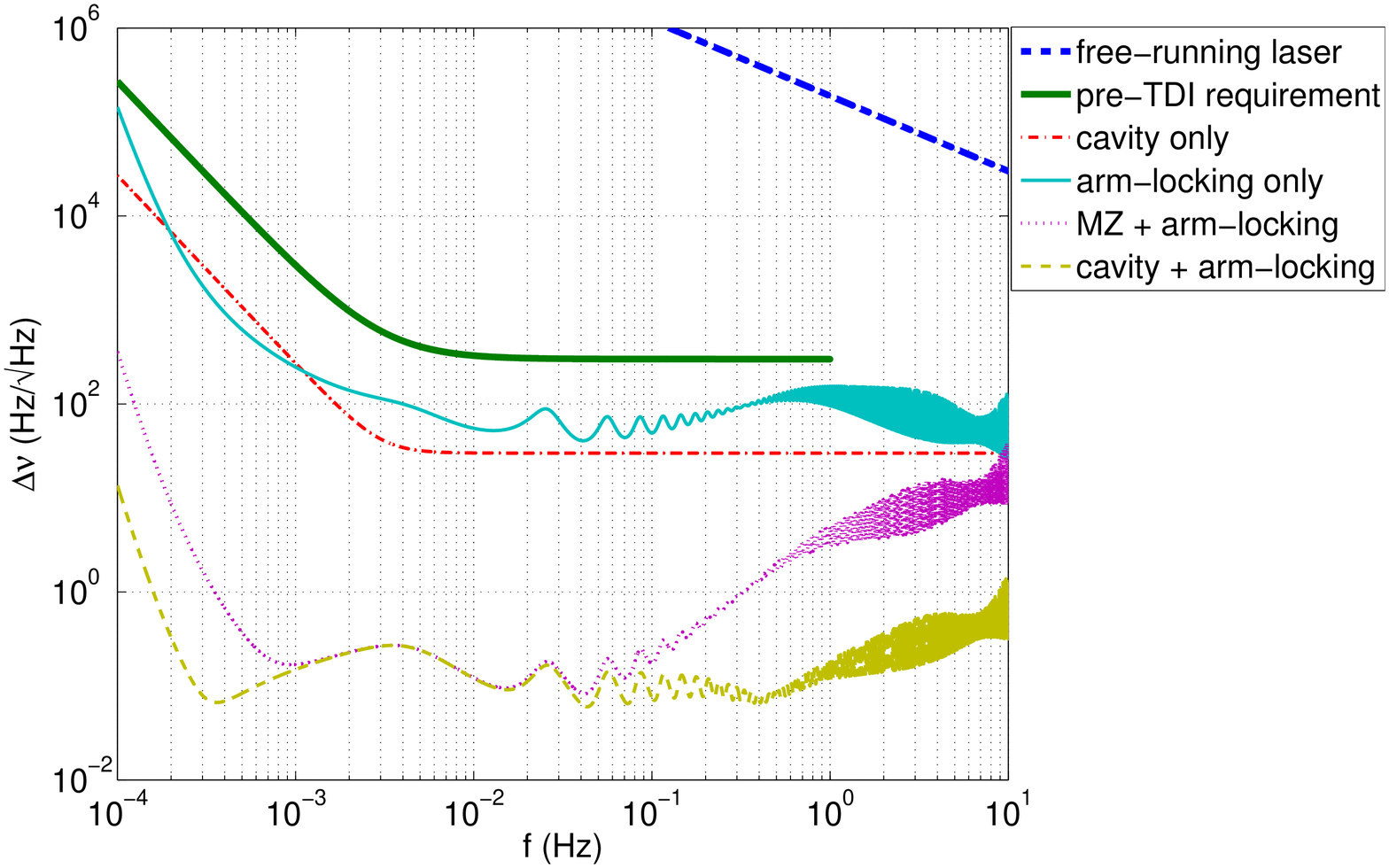}
\par\end{centering}

\caption{\label{fig:StabOptions}Possible options for laser phase noise suppression
in LISA. See text for details.}

\end{figure}

Figure \ref{fig:StabOptions} illustrates the noise performance of
the four reference designs, which are described below.
\begin{itemize}
\item Fixed Cavity (thin dot-dashed line): Master laser locked to a fixed
Fabry-Perot cavity using standard techniques.
\item Arm-locking only (thin solid line): Master laser locked using modified
dual arm locking.
\item Arm-locking with MZ pre-stabilization (thin dotted line): Master laser
pre-stabilized to MZ interferometer with offset. Additional stabilization
provided by modified dual arm-locking.
\item Arm-locking with cavity pre-stabilization (thin dashed line): Master
laser pre-stabilized to fixed Fabry-Perot cavity with an offset using
sideband locking technique. Additional stabilization provided by modified
dual arm locking.
\end{itemize}
For each design, the noise performance lies below the pre-TDI limit (thick
solid line). The selection of a particular design for LISA will be based on criteria other than noise performance such as complexity, cost, easy of ground
testing and integration, etc.

\section{Summary}

The measurement of $10^{-11}\,\mbox{m}$ displacements over $5\times10^{9}\,\mbox{m}$
baselines is one of the technological capabilities that will allow
LISA to measure gravitational waves. Achieving this capability will
require overcoming a number of challenges. Thanks to the work of a large number of researchers over many years,
a design for the LISA long-arm interferometer that meets these challenges
has been developed.

\section*{Acknowledgments}

In addition to recognizing the contributions of the
community as a whole, I would like to thank Kirk McKenzie for providing
the tools needed to compute arm-locking performance, Steve Hughes
for providing LISA orbital data, and Bob Spero for helpful discussions
on TDI limitations.
\par
Copyright (c) 2010 United States Government as represented by the
Administrator of the National Aeronautics and Space Administration. No
copyright is claimed in the United States under Title 17, U.S. Code.
All other rights reserved.

\section*{References}

\bibliographystyle{unsrt}
\bibliography{bibliography}

\end{document}